\documentclass[11pt]{article}

\usepackage{palatino}
\usepackage{mathpazo}

\usepackage{braket}
\usepackage{amsfonts}
\usepackage{amssymb}
\usepackage{amsmath}
\usepackage{latexsym}
\usepackage{amsthm}
\usepackage{eepic}
\usepackage{sectsty}
\usepackage[usenames]{color}
\usepackage{hyperref}
\usepackage{gastex}

\hypersetup{pdfpagemode=UseNone}

\newtheorem{theorem}{Theorem}
\newtheorem{lemma}[theorem]{Lemma}
\newtheorem{prop}[theorem]{Proposition}
\newtheorem{cor}[theorem]{Corollary}
\theoremstyle{definition}
\newtheorem{definition}[theorem]{Definition}

\newcommand{\tinyspace}{\mspace{1mu}}

\newcommand{\op}[1]{\operatorname{#1}}
\newcommand{\abs}[1]{\left\lvert\tinyspace #1 \tinyspace\right\rvert}
\newcommand{\ceil}[1]{\left\lceil #1 \right\rceil}

\newcommand{\norm}[1]{\left\lVert\tinyspace #1 \tinyspace\right\rVert}

\newcommand{\tr}{\operatorname{Tr}}

\def\I{\mathbb{1}}

\def\complex{\mathbb{C}}
\def\real{\mathbb{R}}

\newenvironment{mylist}[1]{\begin{list}{}{
	\setlength{\leftmargin}{#1}
	\setlength{\rightmargin}{0mm}
	\setlength{\labelsep}{2mm}
	\setlength{\labelwidth}{8mm}
	\setlength{\itemsep}{0mm}}}
	{\end{list}}
\newcommand{\class}[1]{\textup{#1}}

\def\A{\mathcal{A}}
\def\B{\mathcal{B}}

\begin{document}

\title{\LARGE\bf Consequences and Limits of Nonlocal Strategies}

\author{
  Richard Cleve\thanks{%
    Institute for Quantum Computing and School of Computer Science,
    University of Waterloo, Canada}
  \and
  Peter H{\o}yer\thanks{%
    Institute for Quantum Information Science and Department of
    Computer Science, University of Calgary, Canada}
  \quad\quad
  Ben Toner\thanks{%
    BQP Solutions Pty Ltd, Elsternwick, Australia}
  \quad\quad
  John Watrous\thanks{%
    Institute for Quantum Computing and School of Computer Science,
    University of Waterloo, Canada}
}

\date{January 11, 2010}

\maketitle

\begin{abstract}
  This paper investigates the powers and limitations of quantum
  entanglement in the context of cooperative games of incomplete
  information. 
  We give several examples of such \emph{nonlocal games} where
  strategies that make use of entanglement outperform all possible
  classical strategies.
  One implication of these examples is that entanglement can
  profoundly affect the soundness property of two-prover interactive
  proof systems.
  We then establish limits on the probability with which strategies
  making use of entanglement can win restricted types of nonlocal
  games.
  These upper bounds may be regarded as generalizations of
  Tsirelson-type inequalities, which place bounds on the extent to
  which quantum information can allow for the violation of Bell
  inequalities.
  We also investigate the amount of entanglement required by optimal
  and nearly optimal quantum strategies for some games.
\end{abstract}

\section{Introduction}
\label{sec:introduction}

This paper studies the implications of quantum entanglement to
\emph{nonlocal games}, which are cooperative games of incomplete
information.
There are two cooperating players in a nonlocal game: \emph{Alice} and
\emph{Bob}. 
A \emph{referee}, also sometimes called a \emph{verifier} in this context,
determines the game as follows.
First, the referee randomly chooses questions for the players, drawn
from finite sets according to some fixed probability distribution, and
sends each player their question.
Without communicating, and therefore without knowing (in general) what
question was asked of the other player, Alice and Bob must each
respond to the referee with an answer.
The referee then evaluates some predicate on the questions asked of
the players together with their answers, to determine whether they win
or lose.

When analyzing a given nonlocal game, we assume that Alice and Bob
have complete knowledge of the probability distribution that
determines the referee's questions, as well as the predicate that
determines whether they win or lose. 
Based on this information they can agree before the game starts on
a joint strategy.
A \emph{quantum strategy} allows Alice and Bob to make use of a shared
entangled state, while a \emph{classical strategy} does not.
For some nonlocal games, quantum strategies can allow the players to
win with a higher probability than is possible for any classical
strategy, due to the fact that measurements of an entangled state can
result in correlations that are not achievable without the use of
entanglement.

Nonlocal games are natural abstractions of multi-prover interactive
proof systems (consisting of a single round of interaction with two
provers), and have previously been considered in this context in
purely classical terms \cite{FeigeL92, CaiCL94, Raz98}.
They also provide a natural setting in which the quantum-physical
notion of \emph{nonlocality} can be cast.
In quantum information theory, a \emph{Bell inequality} is analogous
to an upper bound on the probability with which Alice and Bob can
win a nonlocal game using a classical strategy, and a \emph{violation}
of a Bell inequality is analogous to the situation in which a quantum
strategy wins a particular nonlocal game with a higher probability
than is possible for any classical strategy.
\emph{Tsirelson-type inequalities} are bounds on the amount by which
quantum information can allow for the violation of Bell inequalities,
and therefore are analogous to upper bounds on the probabilities with
which nonlocal games can be won using quantum strategies.

We will begin our study of nonlocal games in
Section~\ref{sec:definitions}, where we give precise definitions of
nonlocal games and the classical and quantum values of these games.
These values represent the optimal probabilities with which Alice and
Bob can win the games using classical and quantum strategies,
respectively.

Then, in Section~\ref{sec:examples}, we present four examples of
nonlocal games for which quantum strategies outperform classical
strategies, including nonlocal games for which there exist
\emph{perfect} quantum strategies (meaning that the strategies win
with probability one), but for which there do not exist perfect
classical strategies.
These examples are not new, but for the most part have been presented
in the theoretical physics literature as hypothetical thought
experiments, and their connections with nonlocal games and
multi-prover interactive proofs are obscure.
The simplicity of some of our presentations may help to elucidate some
of the features of nonlocality.

Next, in Section~\ref{sec:interactive-proofs}, we discuss the
implications of quantum strategies for nonlocal games to the study of
multi-prover interactive proof systems.
In particular, we exhibit two natural examples of two-prover
interactive proof systems that are classically sound, but become
unsound when the provers may employ quantum strategies.

Finally, in Section~\ref{sec:binary}, we provide the beginnings of a
systematic understanding of the limits of nonlocal strategies for two
restricted classes of games: \emph{binary games} and \emph{XOR games}.
Binary games are nonlocal games in which Alice and Bob each respond
with a single bit, and XOR games are binary games for which the
referee's predicate that determines whether Alice and Bob win or lose
depends only on the parity (or exclusive-OR) of their answer bits.
The results proved in this section include generalizations of Tsirelson's 
inequality.
We also prove upper bounds on the amount of entanglement needed to play XOR
games optimally or nearly optimally.

\subsection*{Remark on a previous version of this paper}

A preliminary version of this paper appeared at the 
\emph{IEEE Conference on Computational Complexity} in 2004, and was
posted to the arXiv.org e-print server in April of 2004.
This preliminary version contains an error: one of the technical
lemmas (numbered Lemma 5.4 in the proceedings version and Lemma 5.5 in
the arXiv.org version) is false.
The present version corrects this mistake, which requires a new proof
of Theorem~\ref{thm:perfect-strategies} below.

Aside from this correction, the present version of this paper is
similar to the preliminary version.
While some revisions to the writing, notation, and organization of the
paper have been made, and a few other minor points have been added,
there is (for the most part) no discussion of newer work that was done
subsequent to the writing of the preliminary version.
(Readers interested in further developments on nonlocal games may, for
instance, consult \cite{DohertyLTW08,ItoKPSY08,KempeKMTV08,KempeRT08}.)

\section{Definitions} \label{sec:definitions}

\subsection*{Nonlocal games}

Let $\pi$ be a probability distribution on $S\times T$, and let $V$ be
a predicate on $S\times T\times A\times B$, for finite, non-empty sets
$S$, $T$, $A$, and $B$.
Then $V$ and $\pi$ define a \emph{nonlocal} game $G = G(V,\pi)$ as
follows.
A pair of questions $(s,t)\in S\times T$ is randomly chosen according
to the distribution $\pi$, and $s$ is sent to Alice and $t$ is sent to
Bob.
Alice must respond with an answer $a\in A$ and Bob with an answer $b\in B$.
Alice and Bob are not permitted to communicate after receiving $s$ and
$t$, but they may agree on whatever sort of strategy they like prior to
receiving their questions.
They \emph{win} if $V$ evaluates to~$1$ on $(s,t,a,b)$ and \emph{lose}
otherwise.
To stress the fact that $(a,b)$ is correct or incorrect given questions
$(s,t)$ we will denote the value of the predicate $V$ on $(s,t,a,b)$
as $V(a,b\,|\,s,t)$.

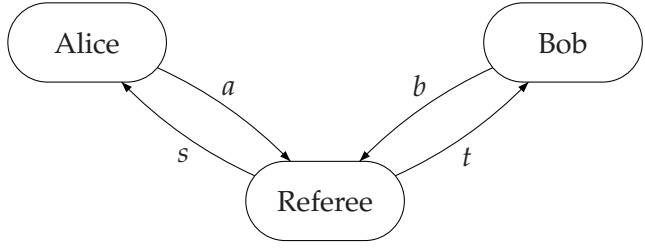
\begin{figure}[t]
\begin{center}
  \unitlength=3pt
  \begin{picture}(80, 30)
    \gasset{Nw=20,Nh=10,Nmr=10}
    \thinlines
    \node(Referee)(40,5){Referee}
    \node(Alice)(10,25){Alice}
    \node(Bob)(70,25){Bob}
    \drawedge[curvedepth=3](Referee,Alice){$s$}
    \drawedge[curvedepth=-3,ELside=r](Referee,Bob){$t$}
    \drawedge[curvedepth=3](Alice,Referee){$a$}
    \drawedge[curvedepth=-3,ELside=r](Bob,Referee){$b$}
  \end{picture}
\end{center}
\caption{The communication structure of a nonlocal game.}
\end{figure}


\subsection*{Classical strategies and classical values of nonlocal
  games}

The \emph{classical value} of a game $G$ is the maximum probability
with which Alice and Bob can win~$G$, ranging over all purely
classical strategies.
The classical value of a game $G$ will be denoted $\omega_c(G)$.
A \emph{deterministic} strategy is a restricted type of classical strategy
in which $a$ and $b$ are simply functions of $s$ and $t$, respectively.
The classical value of a game is always obtained by some deterministic
strategy given that any probabilistic strategy can be expressed as a convex
combination of deterministic strategies.
In other words, it holds that
\[
\omega_c(G(V,\pi)) = \max_{a,b}\sum_{s,t}\pi(s,t)V(a(s),b(t)\,|\,s,t),
\]
where the maximum is over all functions $a:S\rightarrow A$ and
$b:T\rightarrow B$.

\subsection*{Quantum strategies and quantum values of nonlocal games}

We will assume for this discussion and throughout the rest of the paper that
the reader is familiar with the basics of quantum information, which is
discussed (for instance) in the books \cite{NielsenC00},
\cite{KitaevSV02}, and \cite{KayeLM07}.

A quantum strategy for a game $G$ consists of an initial bipartite
state $\ket{\psi}$ shared by Alice and Bob, a quantum measurement for
Alice for each $s\in S$, and a quantum measurement for Bob for each
$t\in T$.
On input $(s,t)$, Alice performs her measurement corresponding to $s$ on 
her portion of $\ket{\psi}$, yielding an outcome $a$.
Similarly, Bob performs his measurement corresponding to $t$ on his 
portion of $\ket{\psi}$, yielding outcome $b$.
The results $a$ and $b$ are sent back to the referee.

In more descriptive mathematical terms, a quantum strategy is
determined by the following items.
\begin{mylist}{\parindent}
\item[1.]
A positive integer $n$ and a unit vector $\ket{\psi} \in \A\otimes\B$,
for $\A$ and $\B$ isomorphic copies of the vector space $\complex^n$.
The space $\A$ represents Alice's part of $\ket{\psi}$ and the
space $\B$ represents Bob's part.

\item[2.]
Two collections of positive semidefinite $n\times n$ matrices
\[
\left\{
X_s^a\,:\rule{0mm}{4mm}\,s\in S,\,a\in A\right\}
\quad\text{and}\quad
\left\{Y_t^b\,:\rule{0mm}{4mm}\,t\in T,\,b\in B\right\}
\]
satisfying
\[
\sum_{a\in A} X_s^a = \I
\quad\text{and}\quad \sum_{b\in B} Y_t^b = \I
\]
for every choice of $s\in S$ and $t\in T$, where $\I$ denotes the
$n\times n$ identity matrix.
For each $s\in S$, the collection $\{X_s^a\,:\,a\in A\}$
describes the measurement performed by Alice on her part of 
$\ket{\psi}$ when she receives the question $s$.
Likewise, for each $t\in T$, the collection $\{Y_t^b\,:\,b\in B\}$
describes Bob's measurement given the question $t$.
\end{mylist}

Given a question $s\in S$ for Alice and a question $t\in T$ for Bob,
such a strategy causes Alice to answer with $a\in A$ and Bob to answer
with $b\in B$ with probability 
$\braket{\psi|X^a_s\otimes Y^b_t|\psi}$.
The probability that Alice and Bob win the game $G$ using such a
strategy is therefore given by
\[
\sum_{(s,t)\in S\times T}\pi(s,t)
\sum_{(a,b)\in A\times B}
\braket{\psi | X^a_s\otimes Y^b_t | \psi}
V(a,b\,|\,s,t).
\]

The \emph{quantum value} of a game $G$, denoted $\omega_q(G)$, 
is the supremum of the winning probabilities over all quantum
strategies of Alice and Bob.
It is not known if the quantum value is always achieved by some
strategy, due to the fact that the number $n$ is not \emph{a priori}
bounded.
For instance, it has not been ruled out that there could sometimes
exist a sequence of strategies requiring increasing values of $n$ that
win a game $G$ with probabilities converging to, but never reaching,
its quantum value.
(As is shown in \cite{LeungTW08}, this phenomenon indeed occurs in an
extended form of nonlocal games in which the referee can send,
receive, and process quantum information.)

\subsection*{Observables}

When we refer to an \emph{observable} in this paper, we are referring
to a Hermitian matrix that one associates with a projective
measurement whose measurement outcomes are real numbers.
More precisely, suppose that we have a measurement described by 
a collection of projection matrices $\{\Pi_1,\ldots,\Pi_k\}$ for which
$\Pi_1+\cdots+\Pi_k = \I$, and suppose further that we associate the
outcomes of the measurement with a collection of real numbers
$\{\lambda_1,\ldots,\lambda_k\}$.
Then the observable corresponding to this measurement is given by
\begin{equation} \label{eq:spectral-decomposition}
A = \sum_{j = 1}^k \lambda_j \Pi_j.
\end{equation}
Given such a matrix $A$, one may determine the corresponding
projective measurement by computing the spectral decomposition of
$A$.

We will primarily use observables when discussing strategies for games
that require Alice and Bob to perform two-outcome measurements,
especially in the case of \emph{binary games} (which are the main
topic of Section~\ref{sec:binary}).
In this setting it is convenient to associate real numbers in the set
$\{+1,-1\}$ with the binary values $\{0,1\}$ using the correspondence
$0 \mapsto +1$ and $1 \mapsto -1$ (as is common when using
discrete Fourier analysis of Boolean functions).
When using this convention, the observable $A$ corresponding to a
binary-valued projective measurement $\{\Pi_0,\Pi_1\}$ is given by
$A = \Pi_0 - \Pi_1$.

\section{Examples of nonlocal games} \label{sec:examples}

The fact that entanglement can cause non-classical correlations is 
a familiar idea in quantum physics, introduced in a seminal 1964 paper 
by Bell~\cite{Bell64}.
In the following subsections, we give four examples of this phenomenon.

\subsection{The CHSH game}\label{sec:chsh}

Our first example of a game for which a quantum strategy outperforms any
classical strategy is a well-known example based on the
\emph{CHSH inequality}, named for its discoverers Clauser, Horne,
Shimony, and Holt \cite{ClauserHSH69}.
Rephrased in terms of nonlocal games, the example is as follows.
Let $S = T = A = B = \{0,1\}$, let $\pi$ be the uniform distribution on
$S\times T$, and let $V$ be the predicate
\[
V(a,b\,|\,s,t) = \left\{
\begin{array}{ll}
1 & \text{if $a\oplus b = s\wedge t$}\\
0 & \text{otherwise.}
\end{array}
\right.
\]
The classical value of the game $G = G(V,\pi)$ is $\omega_c(G) = 3/4$, which
is easily verified by considering all deterministic strategies.
Using a quantum strategy, however, Alice and Bob can win this game 
with probability $\cos^2(\pi/8) \approx 0.85$.
This probability is optimal, so we have $\omega_q(G) = \cos^2(\pi/8)$.
A description of a quantum strategy that achieves this probability 
of success follows, and the fact that it is optimal follows from
Tsirelson's Inequality~\cite{Khalfin85,Tsirelson80}.

First, let the entangled state shared by Alice and Bob be
$\ket{\psi} = (\ket{00}+\ket{11})/\sqrt{2}$, define 
\begin{align*}
\ket{\phi_0(\theta)} & = \cos(\theta)\ket{0} + \sin(\theta)\ket{1},\\[2mm]
\ket{\phi_1(\theta)} & = -\sin(\theta)\ket{0} + \cos(\theta)\ket{1},
\end{align*}
and let Alice and Bob's measurements be given as 
\begin{align*}
X_0^a & = \ket{\phi_a(0)}\bra{\phi_a(0)},\\[1mm]
X_1^a & = \ket{\phi_a(\pi/4)}\bra{\phi_a(\pi/4)},\\[1mm]
Y_0^b & = \ket{\phi_b(\pi/8)}\bra{\phi_b(\pi/8)},\\[1mm]
Y_1^b & = \ket{\phi_b(-\pi/8)}\bra{\phi_b(-\pi/8)}
\end{align*}
for $a,b\in\{0,1\}$.
Each of these matrices is a rank-one projection matrix, so the
measurements Alice and Bob are making are examples of projective
measurements.
Given our particular choice of $\ket{\psi}$, we have
$\bra{\psi}X\otimes Y\ket{\psi} = 
\frac{1}{2}\tr\left(X^{\mathsf{T}}Y\right)$
for arbitrary matrices $X$ and $Y$.
Thus, as each of the matrices $X_s^a$ and $Y_t^b$ is real and symmetric,
the probability that Alice and Bob answer $(s,t)$ with $(a,b)$ is
$\frac{1}{2}\tr\left(X_s^aY_t^b\right)$.
It is now routine to check that in every case, the correct answer is
given with probability $\cos^2(\pi/8)$ and the incorrect answer with
probability $\sin^2(\pi/8)$.

\subsection{The Odd Cycle game}\label{sec:odd}

For the following game, imagine that Alice and Bob are trying to 
convince the referee that an odd cycle of length $n$ is 2-colorable 
(which it is not, as $n$ is odd).
The referee sends the name of a vertex to each of Alice and Bob 
such that the two vertices are either the same or adjacent.
Alice and Bob each send one of two colors back to the referee.
The referee's requirement is that, when the vertices are the same, 
the two colors should agree, and when the vertices are adjacent, 
the colors should be different.

Formally, let $n\geq 3$ be an odd integer, let $S = T = \mathbb{Z}_n$,
and let $A = B = \{0,1\}$.
Take $\pi$ to be the uniform distribution over the set 
$\{(s,t) \in \mathbb{Z}_n \times \mathbb{Z}_n\,:\,\text{$s=t$ or
$s+1\equiv t\:(\bmod\:n)$}\}$ and let $V$ be defined as 
\[
V(a,b\,|\,s,t) = \left\{
\begin{array}{ll}
1 & \text{if $a \oplus b = [s+1\equiv t\:(\bmod\:n)]$}\\
0 & \text{otherwise.}
\end{array}
\right.
\]
(Here we have written $[s+1\equiv t\:(\bmod\:n)]$ to denote the
Boolean value representing the truth or falsehood of the congruence
$s+1\equiv t\:(\bmod\:n)$.)
This is a variation on a game based on the
\emph{chained Bell inequalities} of Braunstein and Caves
\cite{BraunsteinC90} that generalize the CHSH inequality.
It is also discussed by Vaidman~\cite{Vaidman01}.

It is easy to see that $\omega_c(G) = 1 - 1/2n$ for this game.
Any deterministic strategy must fail for at least one of the possible
pairs $(s,t)$, as an odd cycle cannot be 2-colored, while a strategy
achieving success probability $1-1/2n$ is that Alice and Bob let
$a = s\bmod 2$ and $b = t\bmod 2$.

On the other hand, a quantum strategy can attain a success probability 
quadratically closer to~1.
The following quantum strategy \cite{BraunsteinC90} wins with probability
$\cos^2(\pi/4n) \geq 1 - (\pi/4n)^2$.
The shared state is the same as for the CHSH game:
\[
\ket{\psi} = \frac{1}{\sqrt{2}}(\ket{00}+\ket{11}).
\]
This time we define
\begin{align*}
X_s^a & = \ket{\phi_a(\alpha_s)}\bra{\phi_a(\alpha_s)},\\[1mm]
Y_t^b & = \ket{\phi_b(\beta_t)}\bra{\phi_b(\beta_t)},
\end{align*}
where 
\begin{align*}
\alpha_s & = \left(\frac{\pi}{2} - \frac{\pi}{2n}\right)s + \frac{\pi}{4n},
\\[1mm]
\beta_t & = \left(\frac{\pi}{2} - \frac{\pi}{2n}\right)t,
\end{align*}
and where $\ket{\phi_0(\theta)}$ and $\ket{\phi_1(\theta)}$ are as defined
for the CHSH game.
Given questions $(s,t)$, the probability that Alice and Bob answer the same
bit may be calculated to be
$\cos^2(\alpha_s-\beta_t)$,
which implies they answer different bits with probability
$\sin^2(\alpha_s-\beta_t)$.
In case $s = t$ we have
\[
\alpha_s - \beta_t = \frac{\pi}{4n},
\]
so they answer correctly (i.e., with $a=b$) with probability
$\cos^2(\pi/4n)$.
If $s+1\equiv t\;(\bmod\,n)$, on the other hand, we have
\[
\alpha_s - \beta_t = \frac{\pi}{2} - \frac{\pi}{4n},
\]
so they answer correctly (i.e., with $a\not=b$) with probability
$\sin^2(\pi/2 - \pi/4n) = \cos^2(\pi/4n)$.
Therefore this strategy answers correctly with probability $\cos^2(\pi/4n)$
on every pair of questions.  
This quantum strategy is optimal, as we shall show in
Corollary~\ref{cor:odd} later in the paper.

\subsection{The Magic Square game}\label{sec:magic}

The next game we consider is based on the fact that there does not exist
a $3 \times 3$ binary matrix with the property that each row has even parity
and each column has odd parity.
It is a slight variation of an example presented by Aravind~\cite{Aravind02},
which builds on work by Mermin~\cite{Mermin90,Mermin93}.
The idea is to ask Alice to fill in the values in either a row or a column of
the matrix (randomly selected) and to ask Bob to fill in a single entry of the
matrix, that is randomly chosen among the three entries given to Alice.
The requirement is that the parity conditions are met by Alice's answers
(even for rows, odd for columns) and that Bob's answer is consistent with
Alice's answers.

Formally, let $S = \mathbb{Z}_6$ index the six possible questions to
Alice (three rows plus three columns) and let $T = \mathbb{Z}_9$ index
the nine possible questions to Bob (one for each entry of the matrix).
Let $A = \{0,1\}^3$ and $B = \{0,1\}$.
The predicate $V(a,b\,|\,s,t)$ is defined to take value 1 if and only if 
$a$ has
the appropriate parity (0 for a row and 1 for a column) and the entry of $a$
corresponding to $t$ has value $b$.
The distribution $\pi$ is the uniform distribution over 
$\{(s,t) \in S \times T : \text{entry $t$ is in triple $s$}\}$.
It is not hard to see that $\omega_c(G) = 17/18$ for this game.
It should be noted that, although it is convenient to set $A = \{0,1\}^3$ 
for this game, we could take $A = \{0,1\}^2$, because the third bit of 
Alice's output is determined by the first two bits and the parity 
constraints.

Remarkably, there exists a \emph{perfect} quantum strategy for this
game, meaning that the strategy wins with probability 1.
The particular strategy we will describe is derived from
\cite{Aravind02}, where a slight variant of this game is presented.

This strategy is best described using the notion of observables, which
were discussed at the end of Section~\ref{sec:definitions}.
Consider the following $3\times 3$ matrix of observables, each
corresponding to a measurement of two qubits:
\[
\begin{pmatrix}
\sigma_x\otimes\sigma_y & \sigma_y\otimes\sigma_x &
\sigma_z\otimes\sigma_z\\[1mm]
\sigma_y\otimes\sigma_z & \sigma_z\otimes\sigma_y &
\sigma_x\otimes\sigma_x\\[1mm]
\sigma_z\otimes\sigma_x & \sigma_x\otimes\sigma_z &
\sigma_y\otimes\sigma_y
\end{pmatrix}.
\]
Here, the matrices $\sigma_x$, $\sigma_y$, and $\sigma_z$ are the
(non-identity) Pauli matrices:
\[
\sigma_x = \begin{pmatrix}0 & 1\\ 1 & 0\end{pmatrix},
\quad\quad
\sigma_y = \begin{pmatrix}0 & -i\\ i & 0\end{pmatrix},
\quad\quad
\sigma_z = \begin{pmatrix}1 & 0\\ 0 & -1\end{pmatrix}.
\]
Alice and Bob will share two copies of the state
\[
\ket{\psi^-} = \frac{1}{\sqrt{2}}\ket{01} -
\frac{1}{\sqrt{2}}\ket{10}.
\]
To answer any question they receive about the $3\times 3$ square,
Alice and Bob simply measure their parts of the two shared states with
respect to the corresponding observables in the above matrix.
For example, if Alice is asked to assign values to the first row of
the square, she measures her two qubits with respect to the observable
$\sigma_x\otimes\sigma_y$, then with respect to the observable
$\sigma_y\otimes\sigma_x$, and finally with respect to
$\sigma_z\otimes\sigma_z$, each time obtaining a $\pm1$ outcome
corresponding to a binary value.
Similarly, if Bob is asked to assign a value to the $(1,2)$ entry of
the square, he measures his two qubits with respect to the observable
$\sigma_y\otimes\sigma_x$.

To understand why this strategy works perfectly, a few basic
observations are required.
The first observation is that $\ket{\psi^-}$ is a $-1$ eigenvector of
each of the operators $\sigma_x\otimes\sigma_x$,
$\sigma_y\otimes\sigma_y$, and $\sigma_z\otimes\sigma_z$, which
implies
\begin{equation} \label{eq:Pauli-eigenvector}
\braket{\psi^- | \sigma_x\otimes\sigma_x | \psi^-}
= \braket{\psi^- | \sigma_y\otimes\sigma_y | \psi^-}
= \braket{\psi^- | \sigma_z\otimes\sigma_z | \psi^-}
= -1.
\end{equation}
This implies that Alice and Bob's answers are always in agreement, as
the product of their measurement outcomes (considered as $\pm 1$
values) always equals~1.
For instance, if Alice and Bob both measure their two copies of
$\ket{\psi^-}$ with respect to the observable
$\sigma_y\otimes\sigma_x$, the product of their outcomes will be
\[
\bra{\psi^-}\sigma_y\otimes\sigma_y\ket{\psi^-}
\bra{\psi^-}\sigma_x\otimes\sigma_x\ket{\psi^-} = (-1) (-1) = 1.
\]

The second observation is that the Pauli matrices anti-commute in
pairs:
\begin{equation} \label{eq:Pauli-anti-commute}
\sigma_x \sigma_y = - \sigma_y \sigma_x,
\quad\quad
\sigma_x \sigma_z = - \sigma_z \sigma_x,
\quad\quad
\sigma_y \sigma_z = - \sigma_z \sigma_y.
\end{equation}
This implies that the observables commute within any row or column.
For instance, $\sigma_y\otimes\sigma_x$ and $\sigma_z\otimes\sigma_z$
both belong to the first row, and we have
\[
(\sigma_y \otimes \sigma_x)(\sigma_z \otimes \sigma_z)
= 
(\sigma_y\sigma_z) \otimes (\sigma_x\sigma_z)
=
(-\sigma_z\sigma_y) \otimes (-\sigma_z\sigma_x)
=
(\sigma_z\sigma_y) \otimes (\sigma_z\sigma_x)
=
(\sigma_z \otimes \sigma_z)(\sigma_y \otimes \sigma_x).
\]
This implies that Alice can simultaneously measure the three
observables within whichever row or column she was asked---or
equivalently that the outcomes she obtains do not depend on the
order in which she performs these measurements.

The final observation that is required is that the product of the
observables in each row is equal to $\I\otimes\I$, while the product
of the observables in each column is $-\I\otimes\I$.
This implies that Alice's parity requirements are always met.

\subsection{The Kochen-Specker game}
\label{sec:k-s}

This game is based on the Kochen-Specker Theorem, which can be stated 
as follows.

\begin{theorem}[Kochen and Specker \cite{KochenS67}]
\label{thm:ks}
There exists an explicit set of vectors 
$\{\ket{v_0},\ldots,\ket{v_{m-1}}\}$ in 
$\real^3$ that cannot be $\{0,1\}$-colored so that both of the 
following conditions hold:
\begin{mylist}{\parindent}
\item[1.]
For every orthogonal pair of vectors $\ket{v_i}$ and $\ket{v_j}$, 
they are not both colored 1.
\item[2.]
For every mutually orthogonal triple of vectors $\ket{v_i}$,
$\ket{v_j}$, and $\ket{v_k}$, at least one of them is colored 1.
\end{mylist}
\end{theorem}
The original theorem in \cite{KochenS67} used 117 vectors, but this has 
subsequently been reduced to 31 vectors~\cite{Peres93}.
We will assume that every orthogonal pair of vectors in the set 
is part of an orthogonal triple---which is easily achieved by 
adding a few more vectors to the set---and that the vectors are 
normalized.
Connections between the Kochen-Specker Theorem and nonlocality 
have previously been made in~\cite{HeywoodR83}.

The Kochen-Specker game is defined relative to the above set of 
vectors.
Alice receives a random triple of orthogonal vectors as her input 
and Bob receives a single vector randomly chosen from the triple as 
his input.
Alice outputs a trit indicating which of her three vectors is assigned 
color 1 (implicitly, the other two vectors are assigned color 0).
Bob outputs a bit assigning a color to his vector.
The requirement is that Alice and Bob assign the same color 
to the vector that they receive in common.

It is straightforward to show that the existence of a perfect 
classical strategy for this game would violate the Kochen-Specker 
Theorem, so $\omega_c(G) < 1$ for this game.
On the other hand there is a perfect quantum strategy, using 
entanglement $\ket{\psi} = (\ket{00} + \ket{11} + \ket{22})/\sqrt{3}$.
Alice's projectors (for input $\ket{v_i}, \ket{v_j}, \ket{v_k}$) are 
$\ket{v_i}\bra{v_i},\ \ket{v_j}\bra{v_j},\ \ket{v_k}\bra{v_k}$, 
and Bob's projectors (for input $\ket{v_l}$) are $\ket{v_l}\bra{v_l}$
and $\I - \ket{v_l}\bra{v_l}$.

\section{Connections with multi-prover interactive proof systems}
\label{sec:interactive-proofs}

The two-prover interactive proof system model was defined by
Ben-Or, Goldwasser, Kilian, and Wigderson~\cite{Ben-OrGKW88},
and has been the focus of a great deal of study.
Babai, Fortnow, and Lund \cite{BabaiFL91} proved that every language in NEXP
has a two-prover interactive proof system.
Several refinements to this result were made~\cite{CaiCL90,Feige91,LapidotS91},
leading to a proof by Feige and Lov\'asz \cite{FeigeL92} that a language is in
NEXP if and only if it has a two-prover one-round proof system with perfect
completeness and exponentially small soundness error.

In most work on multi-prover interactive proof systems, the provers
are computationally unbounded, subject to the restriction that they
cannot communicate with each other during the course of the
protocol. 
Because the spirit of the interactive proof system paradigm is to
bound the capabilities of the verifier rather than the prover(s), it
is natural to consider prover strategies that entail sharing entangled
quantum information prior to the execution of the proof system.
Note that such a strategy does not necessarily require the computationally
bounded verifier to manipulate (or know anything about) quantum information.
However, much of the study of multi-prover interactive proof systems 
occurred prior to the mid 1990s, when quantum information was not 
well-known within the theoretical computer science community, and 
quantum strategies were generally not considered.
In fact, the methodologies for analyzing these proof systems usually 
make the implicit assumption that provers are restricted to classical 
strategies.

In this section, we consider what happens when the provers can employ 
quantum strategies.
We do not make any change to the verifier, who remains classical, and all
communication between the verifier and the provers remains classical.%
\footnote{Kobayashi and Matsumoto~\cite{KobayashiM03} consider a related 
but different model, where the provers \emph{and} the verifier manipulate 
quantum information and quantum communication occurs between the verifier 
and the provers.}
A natural question is: What is the expressive power of such proof systems?

Let us write MIP and $\class{MIP}^{\ast}$ to distinguish between 
the cases of no shared entanglement and shared entanglement, respectively.
That is, MIP denotes the class of languages recognized by multi-prover 
interactive proof systems where all communication between the provers and 
verifier is classical and the provers do not share entanglement (as has 
been implicitly assumed in previous contexts).
The definition of $\class{MIP}^{\ast}$ is identical to that of MIP, except 
that the provers may share an arbitrary entangled quantum state at the 
beginning of the protocol.
Furthermore, let $\class{MIP}[k]$ and $\class{MIP}^{\ast}[k]$ 
denote the same classes, but with the number of provers set to $k$.
It is known that $\class{MIP} = \class{MIP}[2] = \class{NEXP}$.
We do not know any relationships between $\class{MIP}^{\ast}$,  
$\class{MIP}^{\ast}[2]$ and $\class{NEXP}$, except the trivial
containment 
$\class{MIP}^{\ast}[2] \subseteq \class{MIP}^{\ast}$.

A \emph{one-round} two-prover interactive proof system is one where 
the interaction is restricted to two stages: a query stage where the 
verifier sends information to the provers, and a response stage 
where the provers send information to the verifier.
Note that such a proof system associates a nonlocal game $G_x$ 
to each string $x$ with the following property.
For all yes-inputs $x$, the value of
$\omega_q(G_x)$ is close to one, and, for all no-inputs $x$, the value
of $\omega_q(G_x)$ is close to zero.

We give two examples of natural two-prover one-round proof systems 
that are classically sound, but become unsound when the provers use 
quantum strategies: one is for languages that express graph chromatic 
numbers and the other is for 3-SAT.
These examples are related to the examples in
Section~\ref{sec:examples}.
We also explain why the existing proofs that equate MIP with NEXP break 
down in terms of their methodology in the case of $\class{MIP}^{\ast}$.
Results in \cite{KobayashiM03} imply that, if the amount of entanglement 
between the provers is polynomially bounded, then any language 
recognized by an $\class{MIP}^{\ast}$ proof system is contained in
NEXP; but without this polynomial restriction, we do not know if
this holds.

\subsection{Graph Coloring proof system}

The Odd Cycle game in Section~\ref{sec:odd} can be regarded as a 
protocol where two provers are trying to convince a verifier that a 
particular graph is two-colorable.
This idea generalizes to any graph $G$ and number of colors $k$.
The verifier asks each prover for the color (among $k$ possibilities) 
of a vertex and requires that the colors be the same whenever each 
prover gets the same vertex and different whenever the provers get 
adjacent vertices.
Formally, the game for $G$ and $k$ is as follows.
Let $S = T = V(G)$, let $A = B = \mathbb{Z}_k$, let 
\[
V(a,b\,|\,s,t) = \left\{
\begin{array}{ll}
1 & \text{if ($s = t$ and $a = b$) or ($(s,t) \in E(G)$ and $a \neq b$)}\\
0 & \text{otherwise,}
\end{array}
\right.
\]
and $\pi$ be the uniform distribution on
$\{(s,s) : s \in V(G)\} \cup \{(s,t) \in E(G)\}$.

If $G$ is $k$-colorable then the provers can satisfy $V$ by basing 
their answers on a valid coloring of~$G$.
Therefore, the value of the associated game is $1$.
If $G$ is not $k$-colorable then, for any classical strategy on the part 
of the provers, there must be an inconsistency for some value of $(s,t)$, 
so the classical value of the associated game is at most 
$1 - 1/(|V(G)|+|E(G)|)$.
The verifier can amplify the difference between the two cases 
($k$-colorable and not $k$-colorable) by repeating this game a 
polynomial number of times (in parallel~\cite{Raz98}).
Thus this is a classical two-prover interactive proof system for 
the language consisting of all $k$-colorable graphs.

This proof system breaks down in the case of entangled provers.
Based on a protocol in \cite{BrassardCT99}, there exists a sequence 
of graphs $G_n$ (where $n$ ranges over all powers of two) with 
the following properties.
First, for any $n$, there is a \emph{perfect} quantum strategy for 
the Graph Coloring proof system with graph $G_n$ and $k=n$ colors.
Second, for sufficiently large $n$, $G_n$ is not $n$-colorable.

For any $n$, $G_n$ is simple to describe: it has vertices $\{0,1\}^n$ 
and two vertices are adjacent if and only if the Hamming distance 
between them is $n/2$.
However, results in \cite{BrassardCT99} show that \emph{there 
exists} an $n$ such that $G_n$ is not $n$-colorable, without 
giving an explicit $n$ for which this holds.
(The proof is based on a related result in \cite{BuhrmanCW98}, 
which makes use of a combinatorial result in \cite{FranklR87}.)
The result is made explicit in \cite{GalliardTW02}, where it is 
shown that $G_{16}$ is not 16-colorable.
Thus, the resulting graph for which the Graph Coloring proof system 
breaks down has $2^{16}$ vertices, and it can be simplified by taking 
only half of its vertices, resulting in a graph of 32,768 vertices.

\subsection{3-SAT proof system}

Next we consider a simple and natural two-prover interactive proof
system for proving that 3-CNF formulas are satisfiable.
The verifier sends the first prover (Alice) a clause and the second
prover (Bob) a variable from that clause.
Alice must provide a satisfying assignment to the variables in that
clause and Bob must assign a value for the variable he receives 
that is consistent with Alice's assignment.

This proof system associates a nonlocal game $G_f$ with
every 3-CNF Boolean formula $f$ over variables $x_0,\ldots,x_{n-1}$
with $m$ clauses $c_0,\ldots, c_{m-1}$.
Specifically, we take
\[
S = \mathbb{Z}_m,\quad
T = \mathbb{Z}_n,\quad
A = \{0,1\}^3,\quad \text{and} \quad
B = \{0,1\}.
\]
For each clause, every $a \in \{0,1\}^3$ induces an assignment to 
each variable that occurs in the clause in a natural way.
The predicate $V(a,b\,|\,s,t)$ takes the value 1 if and only if the
assignment for the variables in $c_s$ induced by $a$ satisfies $c_s$
and is consistent with the assignment $x_t = b$, and the distribution
$\pi$ may be taken to be uniform on
\[
\{(s,t) \in S \times T : \text{clause $c_s$ contains variable $x_t$}\}.
\]

If $f$ is satisfiable then $\omega_c(G_f) = 1$ by the two provers 
returning values corresponding to a specific truth assignment.
If $f$ is unsatisfiable then $\omega_c(G_f) \le 1 - 1/3m$, as 
then at least one of the $3m$ possible $(s,t)$ queries must violate 
the predicate.

However, this proof system breaks down in the case of entangled provers.
Upon seeing the aforementioned counterexample for the Graph Coloring 
proof system, Ambainis~\cite{Ambainis01} showed that a counterexample 
for 3-SAT could be based on it.
Intuitively, the idea is to construct a 3-CNF formula that, for truth 
assignment $x$, expresses the statement ``$x$ is a $k$-coloring of $G$.''
Based on the above counterexample graph with 32,768 vertices (the 
smallest that we are aware of), the resulting 3-SAT formula consists 
of roughly $10^{8}$ clauses.

A simpler counterexample can be based on the Magic Square game in
Section~\ref{sec:magic} that consists of 24 clauses.
We will construct an instance of 3-SAT, where the resulting formula 
is not satisfiable but for which there is a perfect quantum strategy 
for the above two-prover proof system.
Let the variables be $x_{00}$, $x_{01}$, $x_{02}$, $x_{10}$, $x_{11}$,
$x_{12}$, $x_{20}$, $x_{21}$, $x_{22}$, which intuitively correspond 
to a $3 \times 3$ Boolean matrix.
There are six parity conditions in the Magic Square game: each row 
has even parity and each column has odd parity.
Each parity condition can be expressed with four clauses.
For example, for the first row, 
\[
(\overline{x}_{00} \vee \overline{x}_{01} \vee \overline{x}_{02}) \wedge 
(\overline{x}_{00} \vee x_{01} \vee x_{02})
\wedge (x_{00} \vee \overline{x}_{01} \vee x_{02}) \wedge 
(x_{00} \vee x_{01} \vee \overline{x}_{02})
\]
is satisfied if and only if $x_{00} \oplus x_{01} \oplus x_{02} = 0$.
Thus 24 clauses suffice to express all six parity conditions.
This formula is unsatisfiable, but the perfect quantum strategy for 
the Magic Square game in Section~\ref{sec:magic} defeats the 
3-SAT game for this formula with certainty.

\subsection{Oracularization paradigm}

The above example also constitutes a counterexample to a commonly-used
primitive that enables a two-prover system to simulate an
\emph{oracle machine}.
An oracle machine is a one-prover interactive system where the 
prover's responses to a series of questions are required to be 
\emph{non-adaptive}.
Non-adaptive means that when the prover receives a series of 
queries $s_1, s_2, \ldots, s_m$, his response to $s_i$ must 
be a function of $s_i$ alone, not depending on any $s_j$ for $j \neq i$.
There is a simple oracle machine proof system for 3-SAT, where a random 
clause is selected and its three variables are sent as three queries 
to the prover, who must return a value for each one.
The verifier accepts if and only if the responses satisfy the clause.
The prover's success probability is less than one whenever the formula 
is unsatisfiable.

Fortnow, Rompel, and Sipser \cite{FortnowRS94} showed that, with a 
second prover, who is sent a single randomly chosen query from those 
of the first prover, the first prover must behave as an oracle or 
be detected with positive probability.
Nevertheless, the above quantum strategy for the magic square game 
is a counterexample to this result for the case of entangled provers.
Because this is a component in the proof that $\class{MIP} =
\class{NEXP}$, this proof does not carry over to the case of
$\class{MIP}^{\ast}$.

\section{Binary games and XOR games} \label{sec:binary}

In this section we focus our attention on two simple types of games that 
we call \emph{binary games} and \emph{XOR games}.
Binary games are games in which Alice and Bob's answers are bits:
$A = B = \{0,1\}$.
XOR games are binary games that are further restricted in that the value 
of the predicate $V$ may depend only on $a\oplus b$ and not on $a$ and $b$
independently.
The CHSH and Odd Cycle games are examples of XOR games.

We begin by establishing some basic properties of binary games and XOR
games, and then prove upper bounds on the quantum values of these games.
In particular, we prove that any binary game having a perfect quantum
strategy also has a perfect classical strategy, and for XOR games
we obtain quantitatively stronger upper bounds through the use of a
theorem due to Tsirelson.
We then prove upper bounds on the amount of entanglement required for
Alice and Bob to play XOR games optimally and nearly optimally, and
finally point out a further connection between these games and
multi-prover interactive proof systems.

\subsection{Optimality of projective measurements for binary games}

In accordance with the description of quantum strategies in
Section~\ref{sec:definitions}, Alice and Bob's strategy for a binary
game consists of a shared entangled state
$\ket{\psi}\in\A \otimes \B$, for $\A$ and $\B$ isomorphic copies of
the space $\complex^n$, together with a collection of matrices
representing measurements:
$X_s^0$ and $X_s^1$, for each $s\in S$, and
$Y_t^0$ and $Y_t^1$, for each $t\in T$.
In this section we observe that these matrices may always be taken to
be projection matrices, even when restricted to the support of the
vector $\ket{\psi}$.
This fact is easily proved, and in some sense it is more direct than
the well-known fact that general measurements can be simulated by
projective measurements.
(Note also that this simulation requires the use of an ancillary
system, and therefore does not immediately imply the claimed fact.)

\begin{prop} \label{prop:projections}
  Let $G$ be a binary game, and let $\ket{\psi}\in\A\otimes\B$ be any
  fixed state shared by Alice and Bob.
  Then among the set of all strategies for $G$ for which the shared
  entangled state is $\ket{\psi}$, there is an optimal strategy for
  which all of Alice's measurements $\{X_s^0,X_s^1\}$ and Bob's
  measurements $\{Y_t^0,Y_t^1\}$ are projective measurements on $\A$ and
  $\B$, respectively.
\end{prop}

\begin{proof}
Consider any choice of measurements $\{X_s^0,X_s^1\}$ and $\{Y_t^0,Y_t^1\}$.
We will show that these measurements can be replaced by projective measurements
on $\A$ and $\B$ without decreasing the probability that
Alice and Bob win.
Because the set of all possible choices for the measurements
$\{X_s^0,X_s^1\}$ and $\{Y_t^0,Y_t^1\}$ can be represented as a compact
set, the theorem then follows from the fact that there must exist
an optimal choice of measurements.

For each choice of $s\in S$, let
\[
X_s^0 = \sum_{j=1}^n \lambda_{s,j}\ket{\phi_{s,j}}\bra{\phi_{s,j}},
\]
be a spectral decomposition of $X_s^0$.
We will consider the possible strategies for Alice that are obtained by
viewing the vectors $\{\ket{\phi_{s,j}}\,:\,s\in S,\,j=1\ldots,n\}$
as being fixed but the eigenvalues 
$\{\lambda_{s,j}\,:\,s\in S,\,j=1\ldots,n\}$ as being variables.
Because it must be the case that $0 \leq X_s^0 \leq \I$, 
these eigenvalues are subject to the constraint that $\lambda_{s,j}\in [0,1]$
for all $s\in S$ and $j = 1,\ldots, n$.
Also note that because $X_s^1 = \I - X_s^0$, it holds that
\[
X_s^1 = \sum_{j=1}^n \left(1 - \lambda_{s,j}\right)
\ket{\phi_{s,j}}\bra{\phi_{s,j}}.
\]

Now, the probability that Alice and Bob's strategy wins is
\[
  \op{Pr}[\text{Alice and Bob win}]
  = \sum_{s,t}\pi(s,t)\sum_{a,b}
  V(a,b\,|\,s,t)\,\bra{\psi}X_s^a\otimes Y_t^b\ket{\psi},
\]
which may be expressed as an affine function of the variables
$\{\lambda_{s,j}\,:\,s\in S,\,j=1,\ldots,n\}$.
Subject to the constraint that each $\lambda_{s,j} \in [0,1]$, this function
must therefore be maximized for some choice for these variables for which
$\lambda_{s,j}\in\{0,1\}$ for each $s\in S$ and $j = 1,\ldots, n$.
For such a choice of the eigenvalues $\lambda_{s,j}$ we have that each of
Alice's measurements $\{X_s^0,X_s^1\}$ becomes a projective measurement.
Repeating a similar argument for Bob's measurements yields the desired
result.
\end{proof}

\subsection{Perfect strategies for binary games}

We now prove that if $G$ is a binary game for which there exists a
perfect quantum strategy, then $G$ has a perfect classical strategy as
well: $\omega_c(G) = 1$.
This fact depends on the assumption that the game is binary.
The Kochen-Specker game discussed in Section~\ref{sec:definitions},
for instance, is a ternary-binary game (i.e., $A = \{0,1,2\}$ and 
$B = \{0,1\}$) for which there exists a perfect quantum strategy, but
no perfect classical strategy.

\begin{theorem}
\label{thm:perfect-strategies}
Let $G$ be a binary game.
If there exists a quantum strategy for $G$ that wins with probability
1, then $\omega_c(G) = 1$.
\end{theorem}

\begin{proof}
  Assume that a perfect quantum strategy for $G$ is given.
  More specifically, we assume that this strategy uses a shared
  entangled state $\ket{\psi}\in\A\otimes\B$, for $\A$ and $\B$
  isomorphic copies of $\complex^n$, and by
  Proposition~\ref{prop:projections} we may assume that Alice and
  Bob's measurements are projective measurements on $\A$ and $\B$,
  respectively.
  It is therefore possible to describe these measurements by two
  collections of $\pm 1$ observables $\{A_s\,:\,s\in S\}$ and
  $\{B_t\,:\,t\in T\}$ on $\A$ and $\B$, respectively, and for the
  remainder of the proof we will consider for the sake of convenience
  that Alice and Bob's answer sets are given by $\{+1,-1\}$ rather
  than $\{0,1\}$.
  We note that the probability that Alice and Bob respond to a given
  question pair $(s,t)\in S\times T$ with the answers
  $(\alpha,\beta)\in\{+1,-1\}\times \{+1,-1\}$ is given by
  \begin{align*}
    q(\alpha,\beta\,|\,s,t) & = \frac{1}{4}\bra{\psi}
    (\I + \alpha A_s)\otimes (\I + \beta B_t) \ket{\psi}\\
    & = \frac{1}{4}
    + \frac{\alpha}{4}\bra{\psi}A_s\otimes\I\ket{\psi} 
    + \frac{\beta}{4} \bra{\psi}\I\otimes B_t\ket{\psi} 
    + \frac{\alpha\beta}{4} \bra{\psi}A_s\otimes B_t\ket{\psi}.
  \end{align*}
  
  We will now define functions $a:S\rightarrow\{+1,-1\}$ and
  $b:T\rightarrow\{+1,-1\}$ that represent a perfect deterministic
  strategy for Alice and Bob.
  First, fix an orthonormal basis
  $\{\ket{\phi_1},\ldots,\ket{\phi_{n^2}}\}$ of $\A\otimes\B$ such that
  $\ket{\phi_1} = \ket{\psi}$ and where
  $\ket{\phi_2},\ldots,\ket{\phi_{n^2}}$ are chosen arbitrarily (subject
  to the constraint of orthonormality).
  Next, define functions
  $k:S\rightarrow\{1,\ldots,n^2\}$ and
  $\ell:T\rightarrow\{1,\ldots,n^2\}$ as
  \begin{align*}
  k(s) & = \min\left\{j\in\{1,\ldots,n^2\}\,:\,
  \bra{\phi_j}A_s\otimes \I\ket{\psi}\not=0\right\}\\
  \ell(t) & = \min\left\{j\in\{1,\ldots,n^2\}\,:\,
  \bra{\phi_j}\I\otimes B_t\ket{\psi}\not=0\right\}
  \end{align*}
  and also define a function
  $\kappa:\complex\backslash\{0\}\rightarrow\{+1,-1\}$ over the nonzero
  complex numbers as
  \[
  \kappa\left(z\right) = \left\{
  \begin{array}{ll}
    +1 & \text{if $\op{arg}(z)\in[0,\pi)$}\\
    -1 & \text{if $\op{arg}(z)\in[\pi,2\pi)$}.
  \end{array}\right.
  \]
  Finally, define 
  \[
  a(s) = \kappa\left(\bra{\phi_{k(s)}}A_s\otimes \I\ket{\psi}\right)
  \quad\text{and}\quad
  b(t) = \kappa\left(\bra{\phi_{\ell(t)}}\I\otimes B_t\ket{\psi}\right).
  \]
  
  We will now prove that the functions $a$ and $b$ define a perfect
  deterministic strategy for $G$, so that $\omega_c(G) = 1$.
  To do this, we prove that every question pair $(s,t)$ is answered
  with $(a(s),b(t))$ by the given quantum strategy with a positive
  probability: $q(a(s),b(t)\,|\,s,t) > 0$.
  Given that the quantum strategy is perfect, it follows that
  $(a(s),b(t))$ is correct for $(s,t)$ whenever $\pi(s,t)>0$.

  \pagebreak[3]
  
  To prove that $q(a(s),b(t)\,|\,s,t) > 0$ for every choice of $s$ and
  $t$, we first observe the following two facts:
  \begin{enumerate}
  \item
    if $\bra{\psi}A_s\otimes \I\ket{\psi}\not=0$, then
    $a(s)\bra{\psi}A_s\otimes \I\ket{\psi} > 0$, and
  \item
    if $\bra{\psi}\I\otimes B_t\ket{\psi}\not=0$, then
    $b(t)\bra{\psi}\I\otimes B_t\ket{\psi} > 0$.
  \end{enumerate}
  The first fact follows from the observation that 
  $\bra{\psi}A_s\otimes \I\ket{\psi}$ is a real number, and if it is
  nonzero we have $k(s) = 1$, so that 
  $a(s) = \op{sign}(\bra{\psi}A_s\otimes \I\ket{\psi})$.
  The second fact is similar to the first.
  It follows that if either or both of the numbers 
  $\bra{\psi}A_s\otimes \I\ket{\psi}$ or
  $\bra{\psi}\I\otimes B_t\ket{\psi}$ is nonzero, then
  \[
  a(s)\bra{\psi}A_s\otimes\I\ket{\psi}
  + b(t)\bra{\psi}\I\otimes B_t\ket{\psi} > 0.
  \]
  As $\bra{\psi}A_s\otimes B_t\ket{\psi}\geq -1$, it follows that
  \[
  q(a(s),b(t)\,|\,s,t) = 
  \frac{1}{4}
  + \frac{a(s)}{4}\bra{\psi}A_s\otimes\I\ket{\psi} 
  + \frac{b(t)}{4} \bra{\psi}\I\otimes B_t\ket{\psi} 
  + \frac{a(s) b(t)}{4} \bra{\psi}A_s\otimes B_t\ket{\psi} > 0.
  \]
  The quantum strategy therefore results in the answers $(a(s),b(t))$ to
  the question pair $(s,t)$ with a nonzero probability.

  The remaining case to consider is that
  $\bra{\psi}A_s\otimes \I\ket{\psi} 
  = \bra{\psi}\I\otimes B_t\ket{\psi} = 0$.
  In this case we have
  \[
  q(a(s),b(t)\,|\,s,t) = \frac{1}{4} + \frac{a(s) b(t)}{4}
  \bra{\psi}A_s\otimes B_t\ket{\psi}.
  \]
  As in the first case, we wish to prove that $q(a(s),b(t)\,|\,s,t)$ is
  positive, so we assume toward contradiction that
  $q(a(s),b(t)\,|\,s,t) = 0$, which implies that
  $a(s)b(t)\bra{\psi}A_s\otimes B_t\ket{\psi}=-1$.
  We have that
  \[
  a(s) b(t)\bra{\psi}A_s\otimes B_t\ket{\psi}
  = \sum_{j = 1}^{n^2}
  a(s) b(t)\bra{\psi}A_s\otimes \I\ket{\phi_j} 
  \bra{\phi_j}\I\otimes B_t\ket{\psi},
  \]
  and we observe that this quantity corresponds
  to the inner product of the two $n^2$ dimensional unit vectors whose
  $j$-th entries are given by $a(s) \bra{\phi_j}A_s\otimes
  \I\ket{\psi}$ and $b(t)\bra{\phi_j}\I\otimes B_t\ket{\psi}$,
  respectively.
  If it is the case that this inner product is $-1$, then it must hold
  that
  \begin{equation} \label{eq:contradiction}
    a(s) \bra{\phi_j}A_s\otimes \I\ket{\psi} 
  = -b(t) \bra{\phi_j}\I\otimes B_t\ket{\psi}
  \end{equation}
  for every choice of $j=1,\ldots,n^2$.
  However, if it is the case that $k(s) \not= \ell(t)$, then
  \eqref{eq:contradiction} must fail to hold for 
  $j = \min\{k(s),\ell(t)\}$, while if $k(s) = \ell(t)$, then for
  $j = k(s) = \ell(t)$ we find that the values
  $\op{arg}(a(s)\bra{\phi_j}A_s\otimes \I\ket{\psi})$ and
  $\op{arg}(b(t)\bra{\phi_j}\I\otimes B_t\ket{\psi})$ are both in the
  range $[0,\pi)$, and therefore \eqref{eq:contradiction} again fails
  to hold.
  We have obtained a contradiction, which completes the proof.
\end{proof}

\subsection{Bounds on values of XOR games} \label{sec:XOR}

We now prove upper bounds on quantum strategies for XOR games, which
are binary games where the predicate $V(a,b\,|\,s,t)$ depends only on
$c = a\oplus b$ and not $a$ and $b$ independently.
It will be convenient to view the predicate $V$ as taking only three inputs
in this case, so we write $V(a \oplus b\,|\,s,t)$ rather than
$V(a,b\,|\,s,t)$.

\subsubsection*{Tsirelson's correspondence}

To establish upper bounds on quantum strategies for XOR games, we
begin with a theorem due to Tsirelson~\cite{Tsirelson87}.
This theorem is most naturally stated in terms of \emph{observables},
as described in Section~\ref{sec:definitions}.
(By Proposition~\ref{prop:projections}, there will be no loss of
generality in considering only projective measurements, or
equivalently their corresponding observables, in the context of binary
games.)

\begin{theorem}[Tsirelson \cite{Tsirelson87}]
\label{theorem:Tsirelson}
Let $S$ and $T$ be finite, nonempty sets, and let
$\{c_{s,t}\,:\,(s,t)\in S\times T\}$ be a collection of real
numbers in the range $[-1,1]$.
Then the following are equivalent:
\begin{mylist}{\parindent}
\item[1.]
There exists a positive integer $n$, complex Hilbert spaces $\A$ and
$\B$ with finite dimension $n$, a unit vector
$\ket{\psi}\in\A\otimes\B$, a collection $\{A_s\,:\,s\in S\}$ of
$\pm 1$ observables on $\A$, and a collection $\{B_t\,:\,t\in T\}$ of
$\pm 1$ observables on $\B$, such that
\[
\bra \psi A_s \otimes B_t \ket \psi = c_{s,t}
\]
for all $(s,t)\in S\times T$.

\item[2.]
There exists a positive integer $m$ and two collections
$\{\ket{u_s}\,:\,s\in S\}$ and $\{\ket{v_t}\,:\,t\in T\}$ of unit
vectors in $\real^m$ such that
\[
\langle u_s | v_{t}\rangle = c_{s,t}
\]
for all $(s,t)\in S\times T$.
\end{mylist}
Moreover, if the first item holds for a fixed choice of $n$, then the
second holds for $m = 2 n^2$; and if the second item holds for a fixed
choice of $m$, then the first holds for $n = 2^{\ceil{m/2}}$.
\end{theorem}

%
%

\subsubsection*{Advantage over the trivial strategy}

Next, to state certain upper bounds on $\omega_q(G)$ for XOR games, it will be 
helpful to define the \emph{trivial random} strategy for Alice and Bob as 
one where they ignore their inputs and answer uniformly generated random 
bits.
If $\tau(G)$ denotes the success probability of game $(G,\pi)$ when 
Alice and Bob are restricted to this trivial strategy, then 
\[
\tau(G) = \frac{1}{2} \sum_{c\in\{0,1\}}\sum_{s,t} \pi(s,t) V(c\,|\,s,t).
\]

\begin{prop}
\label{prop:vectors2}
Let $G(V, \pi)$ be an XOR game and let $m = \min(|S|, |T|)$.
Then
\begin{equation}
\label{eq:maximum}
\omega_q(G) - \tau(G)
= \frac{1}{2} \max_{\{\ket{u_s}\},\{\ket{v_t}\}}
\sum_{s,t} \pi(s,t) \left(V(0\,|\,s,t) - V(1\,|\,s,t)\right)
\braket{u_s | v_t},
\end{equation}
where the maximum is over all choices of unit vectors
$\{\ket{u_s}\,:\,s\in S\}\cup\{\ket{v_t}\,:\,t\in T\}$ in $\real^m$.
\end{prop}

\begin{proof}
Consider any quantum strategy for Alice and Bob given by a shared
entangled state $\ket{\psi}\in\A\otimes\B$ and collections of
observables $\{A_s\,:\,s\in S\}$ and $\{B_t\,:\,t\in T\}$.
We associate with each $A_s$ a real unit vector $\ket{u_s}$ and
with each $B_t$ a real unit vector $\ket{v_t}$, according to
Theorem~\ref{theorem:Tsirelson}.
On input $(s,t)$, the probability that Alice and Bob's answers are
equal is
\[
\braket{\psi | X_s^0 Y_t^0 + X_s^1 Y_t^1 | \psi }
= \frac{1}{2} + \frac{1}{2}\bra{\psi} A_s \otimes B_t \ket{\psi}
= \frac{1}{2} + \frac{1}{2}\braket{u_s | v_t}.
\]
It follows that their answers are not equal with probability 
$\left(1-\braket{u_s|v_t}\right)/2$.
Hence the probability that Alice and Bob win using this strategy is  
\[
\frac{1}{2} \sum_{s,t,c} \pi(s,t) V(c\,|\,s,t)
+ \frac{1}{2}  \sum_{s,t} \pi(s,t)
\left( V(0\,|\,s,t) - V(1\,|\,s,t)\right) \braket{u_s|v_t}
\]

Assuming the spaces $\A$ and $\B$ have dimension $n$, the vectors
$\ket{u_s}$ and $\ket{v_t}$ are unit vectors in $\real^{2n^2}$.
Although the dimension $n$ is \emph{a priori} unbounded, the winning
probability depends only on the inner products among the unit vectors
$\{\ket{u_s}\,:\,s\in S\}$ and $\{\ket{v_t}\,:\,t\in T\}$.
We may therefore project onto the span of
$\{\ket{u_s}\,:\,s \in S\}\cup\{\ket{v_t} \,:\,t \in T\}$, which is a
space with dimension at most $|S| + |T|$. 
Indeed, it is sufficient to project the vectors 
$\{\ket{u_s} \,:\,s \in S\}$ onto the span of the vectors 
$\{\ket{v_t} \,:\,t \in T\}$ (or vice versa).
The dimension of this space is at most $m = \min(|S|,|T|)$.
Without loss of generality, let us assume $|S|\leq |T|$. 
Although the vectors $\{\ket{u_s} \,:\,s \in S\}$ will not necessarily
remain unit vectors after orthogonal projection, the maximum over all
vectors 
$\{\ket{u_s} \in \real^{|T|}\,:\,s\in S,\, \|\ket{u_s}\|\leq 1\}$ is 
achieved by points on the boundary---unit vectors---and so it is
sufficient to restrict to this case. 

We now show this strategy can be realized as a quantum protocol.
The maximization in \eqref{eq:maximum} is over a compact set, so the
maximum is achieved by some choice of vectors
$\{\ket{u_s}\,:\,s\in S\}$ and $\{\ket{v_t}\,:\,t\in T\}$
in $\real^m$.
Let $\ket{\psi}$ be a maximally entangled state on $\lceil m/2 \rceil$
qubits.
By Theorem~\ref{theorem:Tsirelson}, there are observables $\{A_s\}$
and $\{B_t\}$ such that 
\[
\bra \psi A_s \otimes B_t \ket \psi = \braket{u_s|v_t}
\]
for all $s \in S$ and $t \in T$.
Thus the strategy can be realized as a quantum strategy.
\end{proof}

The maximization in Proposition~\ref{prop:vectors2} can be cast as a
semidefinite program, which can be approximated to within an additive 
error of $\varepsilon$ in time polynomial in $|S| + |T|$ and 
in $\log(1/\varepsilon)$. (See Ref.~\cite{Boyd:04} for an introduction 
to semidefinite programming.)

It is trivial to write an expression similar to \eqref{eq:maximum} for
the classical value of an XOR game, viz., 
\[
\omega_c(G) - \tau(G)
= \frac{1}{2} \max_{a(s), b(t)}
\sum_{s,t} \pi(s,t)(V(0\,|\,s,t) - V(1\,|\,s,t))a(s) b(t),
\]
where the maximum is over functions $a:S \to \{-1, +1\}$ and $b: T
\to \{-1, +1\}$.  This integer quadratic program is MAXSNP 
hard~\cite{Alon04}.
Unless P $=$ NP, finding the quantum value of an XOR game is easier than
finding the classical value.

\subsubsection*{Upper bound for XOR games with weak classical strategies}

We now give two bounds for XOR games.
We first consider the regime where the success probability of the 
best classical strategy is not much better than $\tau(G)$, 
the success probability of the trivial random strategy.
In this case no quantum strategy can do significantly better.
The bound will be expressed in terms of Grothendieck's
constant~\cite{Groth53}.
\begin{definition}
Grothendieck's constant $K_G$ is the smallest number such that, for all
integers $N \geq 2$ and all $N \times N$ real matrices $M$, if
\[
\left| \sum_{s,t} M(s,t)\, a_s b_t\right| \leq 1,
\]
for all numbers $a_1,\ldots, a_N$, $b_1,\ldots,b_N$ in $[-1,1]$, then
\[
\left| \sum_{s,t} M(s,t)\, \braket{u_s|v_t} \right| \leq K_G,
\]
for all unit vectors $\ket{u_1},\ldots,\ket{u_N},
\ket{v_1},\ldots,\ket{v_N}$ in $\real^n$ (for any choice of $n$).
\end{definition}

\noindent
(See also \cite{Finch03}.)
Grothendieck's constant is known to satisfy
\[
1.6769 \leq K_G \leq \frac{\pi}{2 \log\left(1 + \sqrt{2}\right)} 
\approx 1.7822,
\]
but the exact value is not known.
The upper bound is due to Krivine~\cite{Krivine:79a} (who conjectures it 
is the exact value), and the lower bound is due to
Davie~\cite{Davie84} and, independently, Reeds~\cite{Reeds91}
(see also~\cite{Fishburn:94a}).

\begin{theorem}
\label{thm:gap}
Let $G$ be an XOR game.  Then 
\begin{equation*}
\omega_q(G)-\tau(G)\leq K_G \left[\omega_c(G) - \tau(G)\right].
\end{equation*}
\end{theorem}

\begin{proof}
Suppose, without loss of generality, that $|S|=|T|$,
and let $N = |S| = |T|$.
Define an $N\times N$ matrix $M$ by
\[
M(s,t) = \frac{1}{2 \left[\omega_c(G) -\tau(G)\right]}\pi(s,t) 
\left[ V(0\,|\,s,t) - V(1\,|\,s,t)\right].
\]
It follows that
\[
\left|\sum_{s,t} M(s,t)\, a_s b_t\right| \leq 1
\]
for all numbers $a_1,\ldots, a_n$, $b_1,\ldots,b_n$ in $[-1,1]$.
By Proposition~\ref{prop:vectors2}, we have
\[
\omega_q(G) - \tau(G)
= \left[\omega_c(G) -\tau(G)\right]
\,\max_{\{\ket{u_s}\},\{\ket{v_t}\}} M(s,t)\, \braket{u_s|v_t}
\leq K_G \left[\omega_c(G) -\tau(G)\right],
\]
which establishes the result.
\end{proof}

For the CHSH game, we have $\tau(G) = 1/2$ and
\[
\omega_q(G) - \tau(G)
=\sqrt{2}\, \left[ \omega_c(G) - \tau(G)\right].
\]
Games for which the ratio of $\omega_q(G) - \tau(G)$ to
$\omega_c(G) - \tau(G)$ is greater than $\sqrt{2}$ can be constructed from
the results in Ref.~\cite{Fishburn:94a}.
In particular, the smallest known game for which
this ratio is larger than $\sqrt 2$ has $|S|=|T|=20$.

\subsubsection*{Upper bound for XOR games with strong classical 
strategies}

We now consider the regime where a classical strategy performs well, 
but not perfectly.
For the Odd Cycle game of Section~\ref{sec:odd}, we obtained 
\[
\omega_c(G) = 1 - \frac{1}{2n}
\]
and
\[
\omega_q(G) \ge \cos^2(\pi/4n)\geq 1 - (\pi/4n)^2;
\]
the quantum strategy is quadratically better than the classical 
one in terms of its failure probability.
In fact such a quadratic improvement is all that is possible for 
XOR games, as will be shown shortly in Theorem~\ref{thm:bound2}.

In order to state and prove Theorem~\ref{thm:bound2}, we first 
define a function $g : [0,1] \rightarrow [0,1]$ and two 
constants, $\gamma_1$ and $\gamma_2$.
The function $g$ has the property that it is minimal subject to 
being concave and bounded below by $\sin^2(\frac{\pi}{2}x)$.
To determine $g$, consider the unique linear mapping 
$h(x) = \gamma_1 x$ such that $h$ is the tangent line to 
$\sin^2(\frac{\pi}{2}x)$ at some point $0 < \gamma_2 < 1$.
It is straightforward to show that 
$g(x) = h(x)$ for $x \le \gamma_2$ and 
$g(x) = \sin^2(\frac{\pi}{2}x)$ for $x > \gamma_2$.
To determine the constants $\gamma_1$ and $\gamma_2$, 
note that the condition on $h$ and the fact that 
$\frac{\text{d}}{\text{d}x}\sin^2(\frac{\pi}{2}x) = 
\frac{\pi}{2}\sin(\pi x)$ imply that 
\begin{equation}
\frac{\pi}{2}\sin(\pi \gamma_2) 
\ = \ \frac{\sin^2(\frac{\pi}{2}\gamma_2)}{\gamma_2}
\ = \ \gamma_1.
\end{equation}

\begin{theorem}
\label{thm:bound2}
Let $G$ be an XOR game with classical value $\omega_c(G)$.  Then
$\omega_q(G) \leq g(\omega_c(G))$, where $g$ is as defined above, i.e.,
\begin{equation}
\omega_q(G) \leq 
\left\{
\begin{array}{ll}
\gamma_1 \omega_c(G)  
& \text{if $\omega_c(G) \leq \gamma_2$}\vspace*{1mm} \\
 \sin^2\left(\frac{\pi}{2}\omega_c(G)\right)
& \text{if $\omega_c(G) > \gamma_2$,} \\
\end{array}
\right.
\end{equation}
where $\gamma_1 \approx 1.1382$ and $\gamma_2 \approx 0.74202$ are as 
defined above.
\end{theorem}

\begin{proof}
Consider an optimal quantum strategy and let 
\[
\{\ket{u_s}\,:\,s \in S\},\;\{\ket{v_t}\,:\, t \in T\}\subset \real^m
\]
be the unit vectors associated with it, according to
Proposition~\ref{prop:vectors2}.  
We use these vectors to define the following classical strategy:
\begin{enumerate}
\item
Alice and Bob share a unit vector $\ket{\lambda} \in \real^m$, chosen
uniformly at random.
\item
When asked question $s$, Alice answers 
$a^{\prime} = \left[1 + \op{sign}(\braket{\lambda | u_s})\right]/{2}$.
\item
When asked question $t$, Bob answers
$b^{\prime} = \left[{1 + \op{sign}(\braket{\lambda | v_t} )}\right]/{2}$.
\end{enumerate}
Here the $\op{sign}$ function is defined by $\op{sign}(x) = +1$ if $x \geq
0$ and $-1$ otherwise. 

Let us calculate the probability that $a^{\prime} \oplus b^{\prime} = 1$.
Introduce an azimuthal coordinate $\phi$ for $\ket{\lambda}$ in the plane 
spanned by $\ket{u_s}$ and $\ket{v_t}$, such that $\ket{u_s}$ has coordinate 
$\phi = 0$ and $\ket{v_t}$ has coordinate 
$\phi = \theta_{st} \equiv \cos^{-1} \braket{u_s|v_t} \in [0,\pi]$.
Then $\op{sign}({\braket{u_s | \lambda}}) = 1$ for $\phi \in [-\pi/2, \pi/2]$ 
and $-1$ otherwise, while $\op{sign}({\braket{v_t | \lambda}}) = 1$ for 
$\phi \in [\theta_{st} - \pi/2, \theta_{st} + \pi/2]$ and $-1$ otherwise.
Because $\ket{\lambda}$ is distributed uniformly in $\real^m$, $\phi$ is 
distributed uniformly in $[0, 2\pi)$.
The probability that $a\oplus b = 1$ is then proportional to the 
measure of the subset of $[0, 2\pi)$ on which 
$\op{sign}({\braket{u_s | \lambda}})= 
  -\op{sign}{(\braket{v_t | \lambda})}$.
In particular, $a' \neq b'$ when 
\[
\phi \in [-\pi/2,\theta_{st}-\pi/2)\cup[\pi/2,\theta_{st}+\pi/2).
\]
Therefore, on input $(s,t)$, 
\begin{equation}
\op{Pr}[a^{\prime} \oplus b^{\prime} = 1] 
= {\textstyle \frac{1}{\pi}}\,\theta_{st}.
\end{equation}
Using the quantum strategy, the probability that 
$a\oplus b = 1$ is given by
\[
\op{Pr}[a \oplus b = 1] 
= {\textstyle \frac{1}{2}}\left(1 - \braket{u_s|v_t}\right) 
= \sin^2\left({\textstyle \frac{1}{2}}\,\theta_{st}\right),
\]
so that  
\[
\op{Pr}[a \oplus b = 1]
= \sin^2\left({\textstyle 
\frac{\pi}{2}}\op{Pr}[a^{\prime} \oplus b^{\prime} = 1]\right) 
\le g(\op{Pr}[a^{\prime} \oplus b^{\prime} = 1]).
\]
Similarly, it can be shown that 
\[
\op{Pr}[a \oplus b = 0] 
= \sin^2\left({\textstyle
 \frac{\pi}{2}}\op{Pr}[a^{\prime} \oplus b^{\prime} = 0]\right) 
\le g\left(\op{Pr}[a^{\prime} \oplus b^{\prime} = 0]\right).
\]

For each $(s,t) \in S \times T$, let $\varpi_c(s,t)$ and $\varpi_q(s,t)$ 
be the probabilities of winning the game when using the classical and
quantum strategies, respectively, given that question $(s,t)$ was asked.
{}From the above, together with the concavity of $g$, it follows that 
$\varpi_q(s,t) \leq g(\varpi_c(s,t))$.
The overall probability of winning using the quantum strategy is 
\[
\sum_{s,t} \pi(s,t) \varpi_q(s,t) 
\le \sum_{s,t}\pi(s,t)  g(\varpi_c(s,t)) 
\le g\left(\sum_{s,t}\pi(s,t) \varpi_c(s,t)\right) 
\le g(\omega_c(G)),
\]
where we have again used the fact that $g$ is concave.
\end{proof}

We emphasize that our means of defining the classical strategy in the 
above proof is not 
original; indeed we can trace the technique back to Grothendieck, who 
used it to establish the first upper bound on the constant that bears 
his name~\cite{Groth53}.
More recently, Goemans and Williamson used the same idea to derive
randomized approximation algorithms for MAX-CUT and related 
problems~\cite{GoemansW95}.

One consequence of Theorem~\ref{thm:bound2} is that the quantum 
strategy for the Odd Cycle game given in Section~\ref{sec:odd} 
is optimal.
\begin{cor}\label{cor:odd}
For $G$ being the Odd Cycle game, we have $\omega_q(G) = \cos^2(\pi/4n)$.
\end{cor}

\subsection{Bounds on entanglement for XOR games}
\label{sec:xbounds}

\noindent
The final results we prove concern the \emph{amount} of entanglement needed for
Alice and Bob to play a given game optimally.
With respect to this question, our results are restricted to XOR games.
The following theorem follows immediately from the results of
Section~\ref{sec:XOR}.

\begin{theorem}
Let $G$ be an XOR game and let $m = \min (|S|,|T|)$.
There exists an optimal strategy for Alice and Bob for $G$ in which they
share a maximally-entangled state on $\lceil m/2 \rceil$ qubits.
\end{theorem}

\noindent
Unfortunately, even in this restricted setting of XOR games, the
bound on the amount of entanglement provided by this theorem is
still huge---the number of qubits shared by Alice and Bob is exponential
in the sizes of their inputs.

However, if we are willing to settle for a slightly sub-optimal strategy,
a polynomial number of shared qubits suffices.
This fact follows from the Johnson-Lindenstrauss lemma~\cite{Johnson:84a},
which we now state, following Ref.~\cite{Dasgupta:99a}.

\begin{lemma}[Johnson-Lindenstrauss]
For $\varepsilon\in(0,1)$ and $n$ a positive integer, let $K$ be a positive 
integer
such that $$ K \geq 4 \left( \varepsilon^2/2 - \varepsilon^3/3\right)^{-1} 
\log n.$$
Then for any set $V$ of $n$ points in $\real^d$, there is a mapping
$f:\real^d \to \real^K$ such that for all $\ket{u}, \ket{v}\in V$, 
\[
(1-\varepsilon) \|\ket{u}-\ket{v}\|^2 
\leq \|f(\ket{u})-f(\ket{v})\|^2 \leq (1+\varepsilon) 
\|\ket{u}-\ket{v}\|^2.
\]
\end{lemma}

\begin{theorem}\label{thm:entanglement}
Let $G=G(V, \pi)$ be an XOR game with quantum value $\omega_q(G)$.
Let $0 < \varepsilon < 1/10$, and suppose $K$ is an even integer such that
\[
K \geq 4 \left( \varepsilon^2/2 - \varepsilon^3/3\right)^{-1}
\log\left(|S|+|T|+1\right).
\]
Then, if Alice and Bob share a maximally entangled state
on $K/2$ qubits, they can win with probability greater
than $\omega_q(G) - \varepsilon$.
\end{theorem}

\begin{proof}
Let $M = |S|+ |T|$ and $0 < \varepsilon < 1/10$.
Let $\left\{\ket{u_s}\,:\,s\in S\right\}$ and
$\left\{\ket{v_t}\,:\,t\in T\right\}$
be the vectors associated with the optimal quantum strategy according
to Proposition~\ref{prop:vectors2}.
Let $f$ be the mapping obtained from the Johnson-Lindenstrauss Lemma
to the $M+1$ points
\[
\{\ket{u_s}\,:\,s\in S\},\quad\{\ket{v_t}\,:\,t\in T\}, \quad
\text{and}\quad 0.
\]
Set
\[
\ket{u_s'} = \frac{f(\ket{u_s}) - f(0)}{\|f(\ket{u_s}) - 
f(0)\|}
\quad\text{and}\quad
\ket{v_t'} = \frac{f(\ket{v_t}) - f(0)}{\|f(\ket{v_t}) - f(0)\|}.
\]
for each $s \in S$ and $t\in T$.
Because $\braket{x|y} = 1 - \|x - y \|/2$ for real unit vectors $x$
and $y$, we have 
\[
\left|\,\braket{u_s'| v_t'} - \braket{u_s|v_t}\,\right| = 
\frac{1}{2} \left|\,\|\ket{u_s'} - \ket{v_t'}\| - 
\|\ket{u_s} - \ket{v_t}\|\,\right|.
\]
A straightforward calculation 
based on the fact that
\begin{align*}
\sqrt{1 - \varepsilon} & \leq
\| f(\ket{u_s}) - f(0) \|
\leq \sqrt{1 + \varepsilon},\\[2mm]
\sqrt{1 - \varepsilon} & \leq
\,\| f(\ket{v_t}) - f(0) \|
\leq \sqrt{1 + \varepsilon},
\end{align*}
and 
\[
\sqrt{1 - \varepsilon} \|\ket{u_s} - \ket{v_t}\|
\leq \| f(\ket{u_s}) - f(\ket{v_t}) \| \leq \sqrt{1 + \varepsilon}
\|\ket{u_s} - \ket{v_t}\|
\]
proves that
\[
\frac{1}{2}
\abs{\norm{\ket{u_s'} - \ket{v_t'}} - \norm{\ket{u_s} - \ket{v_t}}}
< 2\varepsilon.
\]

We note that these vectors can be realized as a quantum strategy by 
Theorem~\ref{theorem:Tsirelson}.
It follows that the difference in the probability of winning using 
this strategy instead of the optimal one is 
\begin{align*}
\frac{1}{2} \sum_{s,t} \pi(s,t)\left[V(0\,|\,s,t) - V(1\,|\,s,t)\right]
\left(\braket{u_s|v_t} -\braket{u_s'|v_t'}\right)\hspace{-7cm}\\
& \leq
\frac{1}{2} \sum_{s,t} \pi(s,t) \left|\braket{u_s|v_t} -
\braket{u_s'|v_t'}\right|
\leq {\epsilon} \sum_{s,t} \pi(s,t) = {\varepsilon}.
\end{align*}
Hence Alice and Bob win using this strategy with probability greater 
than $\omega_q(G) - \varepsilon$.
\end{proof}

Oded Regev has described to us an improved form of this theorem where
$K$ has no dependence on $\abs{S}$ and $\abs{T}$.

\subsection{Further connections with multi-prover interactive proof
  systems}

One motive for considering upper bounds on the quantum values 
of games in general is due to their connections with multi-prover 
interactive proof systems.
For example, recall that the Odd Cycle game can be regarded as a simple 
proof system for the two-colorability of odd cycles---for which the 
correct response of the verifier is to reject.
Although this is valid as a classical two-prover interactive proof system, 
if the quantum value of the game were one (or exponentially close to one) 
then it would not be valid as a quantum proof system.
The upper bound on the value of the Odd Cycle game proved in this 
section (Corollary~\ref{cor:odd}) implies that it is a valid quantum 
proof system, and with a polynomial number of repetitions%
\footnote{In the absence of a quantum analogue of Raz's Parallel 
Repetition Theorem \cite{Raz98}, the repetitions can be applied 
sequentially.}, 
the probability of the verifier incorrectly accepting can be made 
arbitrarily close to zero.
For any one-round two-prover quantum interactive proof system, the 
soundness condition will correspond to a nontrivial upper bound of the 
quantum value of a nonlocality game.
Therefore upper bounds are important tools for analyzing such proof 
systems.

Regarding upper bounds on entanglement required by an optimal 
quantum strategy, we note that results in \cite{KobayashiM03} imply 
that if a polynomial upper bound can be established then 
$\class{MIP}^{\ast}[2] \subseteq \class{NEXP}$.
This indicates that upper bounds on entanglement are also relevant 
for analyzing such proof systems.

\begin{definition}
For $0 \le s < c \le 1$, let 
$\oplus\class{MIP}_{c,s}[2]$ denote the class of all languages $L$ 
recognized by classical two-prover interactive proof systems of the 
following form:
\begin{itemize}
\item
They operate in one round, each prover sends a single bit 
in response to the verifier's question, and the verifier's decision 
is a function of the parity of those two bits.
\item
If $x \not\in L$ then, whatever strategy Alice and Bob follow, the 
Prover's acceptance probability is at most $s$ (the \emph{soundness} 
probability).
\item
If $x \in L$ then there exists a strategy for Alice and Bob for 
which the Prover's acceptance probability is at least~$c$ 
(the \emph{completeness} probability).
\end{itemize}
\end{definition}
\begin{definition}
For $0 \le s < c \le 1$, let 
$\oplus\class{MIP}^{\ast}_{c,s}[2]$ denote the class corresponding to 
the previous definition, where all communication remains classical, 
but where the provers may share prior quantum entanglement.
\end{definition}
The following result is implicit in the work of H{\aa}stad~\cite{Hastad01},
based on the application of methods described in~\cite{BellareGS98}.

\begin{theorem}\label{thm:Hastad}
For all $\varepsilon \in (0,1/16)$, if $s = 11/16 + \varepsilon$ 
and $c = 12/16$ then 
$\oplus\class{MIP}_{c,s}[2] = \class{NEXP}$.
\end{theorem}

\begin{proof}[Sketch of Proof]
We refer the reader to \cite{BellareGS98,Hastad01} for all detailed 
information about probabilistically checkable proof systems (PCPs).
Let $\class{PCP}_{c,s}[r,k]$ denote the class of languages recognized 
by PCPs that makes $k$ queries on the basis of $r$ random bits, and 
have completeness and soundness probabilities $c$ and $s$ respectively.
That is, a verifier can query $k$ bits of a purported proof, 
selected on the basis of $r$ random bits, and makes a determination 
of language membership on the basis of those $k$ values.
A language $L$ is in $\class{PCP}_{c,s}[r,k]$ if: 
(a) for all $x \in L$, there exists a proof for which the verifier's 
acceptance probability is at least $c$; and (b) for all $x \not\in L$, 
the verifier's acceptance never exceeds $s$.
H{\aa}stad~\cite{Hastad01} essentially shows that, for all 
$\varepsilon > 0$, if $s = 11/16 + \varepsilon$, and $c = 12/16$ then 
$\class{PCP}_{c,s}[O(\log n),2] = \class{NP}$ using PCPs where the 
verifier's determination is based on the XOR of the two queried bits.
This can be scaled up one exponential in $n$ along the lines discussed 
in~\cite{BellareGS98} to yield 
$\class{PCP}_{c,s}[n^{O(1)},2] = \class{NEXP}$ with the same XOR property.
Moreover, the proof system has the feature that, if each possible 
pair of queries is taken as an edge of a graph then the resulting 
graph is bipartite.
This means that the PCP can be converted into a two-prover interactive 
proof system with the same completeness and soundness probabilities 
($c$ and $s$) as follows.
The verifier randomly chooses an edge, just as in the PCP, and 
sends one query to Alice and one to Bob, according to the bipartite
structure of the graph.
\end{proof}

An obvious question is: 
Do there exist $c$ and $s$ (with $0 \le s < c \le 1$) 
such that $\oplus\class{MIP}^{\ast}_{c,s}[2] = \class{NEXP}$?
The answer is that this cannot be the case unless $\class{EXP} =
\class{NEXP}$, following from the discussion in Section~\ref{sec:XOR}.

\begin{cor}
For all $s$ and $c$ such that $0 \le s < c \le 1$, 
$\oplus\class{MIP}^{\ast}_{c,s}[2] \subseteq \class{EXP}$.
\end{cor}

\subsection*{Acknowledgments}

\noindent
This work was primarily done while RC and JW were at the University of
Calgary and BT was at the California Institute of Technology, and was
partially supported by Canada's NSERC, MITACS, PIMS, and CIFAR;
the U.S. National Science Foundation under Grant No.~EIA-0086038;
and Alberta's iCORE.

We would like to thank Andris Ambainis, Dave Bacon, Anne Broadbent, 
Andrew Doherty, Nicolas Gisin, David Mermin, Asher Peres, John
Preskill, Oded Regev, Madhu Sudan, and Alain Tapp for helpful
discussions, Steven Finch for providing us with copies of
Refs.~\cite{Davie84} and~\cite{Reeds91}, and the anonymous referees
for several helpful comments and corrections.

\bibliographystyle{plain}
\bibliography{nonlocal-games}

\end{document}